\documentclass[twocolumn,showpacs,amsmath,amssymb,prl]{revtex4}

\usepackage{amsmath,amssymb,latexsym,graphics,epsfig}
\usepackage{pstricks}

\newpsobject{showgrid}{psgrid}{subgriddiv=1}

\begin{document}

\title{Finite-size scaling in extreme statistics }

\author{G. Gy\"{o}rgyi}

\author{N. R. Moloney}

\author{K. Ozog\'{a}ny}

\author{Z. R\'{a}cz}

\affiliation{Institute for Theoretical Physics - HAS,
  E\"{o}tv\"{o}s University, P\'{a}zm\'{a}ny
  s\'{e}t\'{a}ny 1/a, 1117 Budapest, Hungary}
\date{\today}

\begin{abstract}
We study the convergence and shape correction to the limit
distributions of extreme values due to the finite size (FS) of
data sets. A renormalization method is introduced for the case
of independent, identically distributed (iid) variables, showing
that the iid universality classes are subdivided according to the
exponent of the FS convergence, which determines the leading
order FS shape correction function as well. We find that,
for the correlated systems of subcritical percolation and $1/f^\alpha$
stationary ($\alpha<1$) noise, the iid shape correction
compares favorably to simulations. Furthermore, for the strongly correlated
regime ($\alpha>1$) of $1/f^\alpha$ noise, the shape correction
is obtained in terms of the limit distribution itself.
\end{abstract}

\maketitle

Extreme value statistics (EVS) has been much studied in engineering
\cite{Gumbel:1958}, finance \cite{EmrechtETAL:2004} and environmental
sciences \cite{KatzParlangeNaveau:2002} where extreme events may
have disastrous consequences. Recently, interest in EVS has also
been growing in physics, e.g. in spin glasses \cite{MezardETAL:1997},
interface fluctuations \cite{GyorgyiETAL:2003,MajumdarComtet:2004},
and front propagations \cite{KrapivskyMajumdar:2000}.  Unfortunately,
the use of EVS is hampered by the cost of acquiring good quality
statistics: EVS
is derived from the extremes of subsets of a data set, requiring
abundant data for reasonable statistics.  Data analysis is further
complicated by the fact that, while the EVS limit distribution may be
known, the convergence with increasing sample size is slow.  Clearly,
a detailed finite-size (FS) analysis providing the {\em convergence
rate} and {\em shape corrections} to the limit distribution is much
needed.  While for iid variables FS studies exist in the mathematical
literature \cite{DeHaanFerreira:2006}, for correlated systems the
convergence rate and shape corrections are known only in a few cases,
such as Brownian motion \cite{SchehrMajumdar:2006}.

In this Letter we use analytic and phenomenological approaches,
combined with simulations, to investigate FS scaling in EVS.  First,
we develop a renormalization group (RG) method, in which the limit
distribution is a fixed point of the flow in function space of the
finite-sample EVS distributions.  Applied to iid variables, the
approach provides an intuitive and accessible summary of the
mathematical results for the leading FS correction, including the
explicit forms of the shape corrections (scaling functions).  Next, we
consider two systems with correlated variables, namely percolation and
$1/f^{\alpha}$ signals. We numerically study the distribution of the
largest clusters in subcritical percolation.  While the limit
distribution is known to be an iid problem
\cite{Bazant:2000,HofstadRedig:2006}, we find that even the FS
correction fits the iid prediction well. In the case of the maximum
statistics of $1/f^{\alpha}$ signals, $0 \le \alpha <1$ corresponds to
the weakly correlated regime, with an iid limit distribution
\cite{Berman:1964}.  Our simulations indicate that the FS properties
are very close to the iid case for $0 \le \alpha \lesssim 0.5$, but
deviations appear for $0.5 \lesssim \alpha <1$. For $\alpha >1$,
however, the convergence becomes fast (power law) and we
can show that, under a mild assumption, the FS shape correction
is given in terms of the limit distribution and, furthermore, the order
as well as the shape of the correction strongly depends on the way
the distribution is scaled. The paper is concluded by remarks on higher
order FS corrections.

%%%%%%%%%%%%%%%%%%%%%%%%%%%%%%%%%%%%%%%%%%%%%%%%%%%%%%%%

The case of iid variables has been extensively studied
\cite{Galambos:1978}, and we begin our FS study by a reinterpretation
of the original derivation of the extreme limit distributions
\cite{FisherTippett:1928}.  Consider random variables
$z_1,z_2,\dots,z_N$ with parent density $\rho(z)$ and integrated
distribution $\mu(z)=\int_{-\infty}^z\rho(s)ds$. The maximum of the
$z_i$ has the integrated distribution $\mu^N\!(z)$ and the basic
observation \cite{FisherTippett:1928} is that if, after an appropriate
scale change $z=a_Nx+b_N$, $\mu^N\!(z)$ tends to a
limit distribution $M(x)$ as $N\to\infty$, then the same limit should
be reproduced for another $N'=pN$.  This requirement can be
cast in the form
\begin{eqnarray}
 M(x) =[\hat R_p M](x)\equiv  M^p(a_px+b_p)
\label{eq:fp-m}
\end{eqnarray}
where the r.h.s.\ defines $\hat R_p$, which can be interpreted as an RG
operator based on the analogy with critical phenomena.  Indeed, the
operation of raising to power $p>1$ and shifting by $b_p$ eliminates the
irrelevant small argument part of the parent distribution, $a_p$
rescales the relevant ``degrees of freedom", and the fixed point
condition, Eq.~\eqref{eq:fp-m}, determines the limit distribution.

Eq.~\eqref{eq:fp-m} is solved by $a_p=p^\gamma,\ \
b_p=\gamma^{-1}(p^\gamma-1)$ and $M(x)=\exp{[-(1+\gamma
x)^{1/\gamma}]}$, where the final scale of $x$ and the position of the
distribution are set by the standardization $M(0)=M'(0)=1/e$. We thus
have a line of fixed points parameterized by $\gamma$, and $M(x)$ is
the generalized extreme value distribution. The traditional
universality classes are called Fr{\' e}chet (power decay of parent at
infinity), FTG (Fisher-Tippett-Gumbel, faster than power decay), and
Weibull (power decay at a finite cutoff), and correspond to
$\gamma>0,\, =0,\, <0$, respectively.

In the RG picture, the FS behavior is determined by the action of  the
RG transformation on the neighborhood of the fixed point $M(x)$. Thus,
we consider distributions as $M_\epsilon(x) = M(x+\epsilon \psi(x))$,
assuming $\epsilon$ is small. Standardization implies
$\psi(0)=\psi'(0)=0$, and the scale of $\epsilon$ is set by
$\psi''(0)=1$.  Our central observation is that the large $N$ behavior
corresponds to the eigenvalue problem
\begin{equation}
M_{\epsilon'}(x) = [\hat R_p M_{\epsilon}](x) =
M^p_{\epsilon}(a_{p,\epsilon} x+b_{p,\epsilon}),
\label{eq:rg-lin}
\end{equation}
where linearization in $\epsilon$ is understood, $a_{p,\epsilon},
b_{p,\epsilon}$ differ from the fixed point values $a_p, b_p$
determined above to $O(\epsilon)$, and the eigenvalue is
$\lambda=\epsilon'/\epsilon$. Using the fixed point relation
\eqref{eq:fp-m}, we obtain $(\lambda/a_p) \psi''(x) =
\psi''(a_px+b_p)$ whose solution with $p$-independent $\psi(x)$
reads as
\begin{eqnarray}
\psi(x) \!&=&\!\left[ (1\!+\!\gamma x)^{\gamma'\!/\gamma+1}
\!-(\gamma'\!\!+\!\gamma)x \!-\!1\right]
/\gamma'(\gamma'\!\!+\gamma\!)
\label{eq:psi}\\
\lambda&=&p^{\gamma'}.
\label{eq:evalue}
\end{eqnarray}
Thus we see that, for a given universality class parameterized by
$\gamma$, a new parameter $\gamma'$ emerges characterizing the
eigenvalues and eigenfunctions of Eq.~\eqref{eq:rg-lin}. Note that shape
corrections equivalent to $\psi$ have been obtained by direct methods
in the mathematical literature
\cite{DeHaanResnick:1996,DeHaanFerreira:2006}.

In order to link the RG result with the $N$ dependence, we write
$\epsilon=\epsilon_N$ and use \eqref{eq:evalue} to find
$\epsilon'=\epsilon_{pN}=p^{\gamma'} \epsilon_N$.  Assuming a power
form one obtains
\begin{equation}
\epsilon_N\propto N^{\gamma'} \, ,
\label{eq:gammap}
\end{equation}
or, more precisely, $\frac{d\ln|\epsilon_N|}{d\ln N}\to \gamma'$.
Thus $\gamma'$ is the FS convergence rate (stability implies
$\gamma'\leq 0$).  To find $\epsilon_N$ and thus $\gamma'$ for a
given parent, $\mu(z)$, we study the integrated distribution function
$\mu^N(z)$ with the shift and scale parameters $b_N=h(\ln N)$,
$a_N=h'(\ln N)$ expressed through $h(y)=\mu^{-1}(e^{-e^{ -y}})$.
Close to the fixed point one has
\begin{eqnarray}
M_N(x) = \mu^N(a_Nx+b_N)\approx M(x+\epsilon_N\psi(x)),
\end{eqnarray}
and differentiating $-\ln[ -\ln M_N(x)]$ twice at $x=0$ gives, to leading
order, ${da_N }/{db_N}\to \gamma$, and at next order
\begin{equation}
 \epsilon_N= \gamma-{da_N}/{db_N}\,\sim \,N^{\gamma'}.\label{eq:epsN}
\end{equation}
The convergence rate $\gamma'$ is now determined and the perturbation
function $\psi(x)$ follows from \eqref{eq:psi}. This gives a practical
meaning to the results from RG theory.

For data analysis it is convenient to represent the FS correction
with zero mean $\langle x\rangle$ and unit variance $\sigma_x^2$
(finite for $\gamma <1/2$) by using the variable $y=(x-\left\langle
x\right\rangle)/\sigma_x$.  Here
we consider the FTG class ($\gamma=0$) with limit distribution
$M^{(0)}(y)=e^{-e^{-(ay+b)}}$ where $a=\pi/\sqrt{6},\, b=\gamma_E$,
the latter being Euler's constant.  Writing $M_N(y)\approx
M^{(0)}(y)+\epsilon_NM^{(1)}(y)$, we have for the correction
\begin{eqnarray}
 M^{(1)}(y) = \, P^{(0)}(y) \, [\,
 e^{\gamma'(ay+b)} +\alpha y +\beta \, ]\, /\, {a\gamma'^2},
 \label{eq:fs-ftg-gen}
\end{eqnarray}
where $P^{(0)}(y)=M^{(0)}(y)'$,
$\alpha=\Gamma(1-\gamma')\frac{b+\Psi(1-\gamma')}{a}$, with $\Psi(z) =
\Gamma'(z)/\Gamma(z)$ and $\beta =-\Gamma(1-\gamma')$.  For
$\gamma'=0$, Eq.(\ref{eq:fs-ftg-gen}) becomes
[$\zeta(z)$ denotes Riemann's zeta function]
\begin{eqnarray}
M^{(1)}(y) = \, P^{(0)}(y) \, [
\, a^3(y^2-1) - 2\zeta(3)y \, ]\, /\, 2a^2\, .
\label{eq:fs-ftg-null}
\end{eqnarray}

We illustrate the above results on FTG class parents with the commonly
found asymptote $1-\mu(z)\propto e^{-z^\delta}/z^\theta$ ($\delta>0$).
Using Eq.~\eqref{eq:epsN}, for $\delta\neq 1$ we have
$\epsilon_N\approx (\delta-1)/(\delta\ln N)$, so $\gamma'=0$.  For
$\delta=1$, $\theta=0$ (exponential distribution) one finds
$\epsilon_N\approx 1/2N$ and $\gamma'=-1$ while, for $\delta=1$ and
$\theta\neq 0$, we have $\epsilon_N\approx -\theta/\ln^2N$, so again
$\gamma'=0$. Thus generically $\gamma'=0$, with FS shape correction
given by \eqref{eq:fs-ftg-null} and the perturbation decaying
logarithmically.  Faster, $1/N$, convergence is seen only for
$\delta=1$, $\theta=0$ where Eq.~\eqref{eq:fs-ftg-gen} applies with
$\gamma '=-1$.

\begin{figure}
\includegraphics[width=8.3cm]{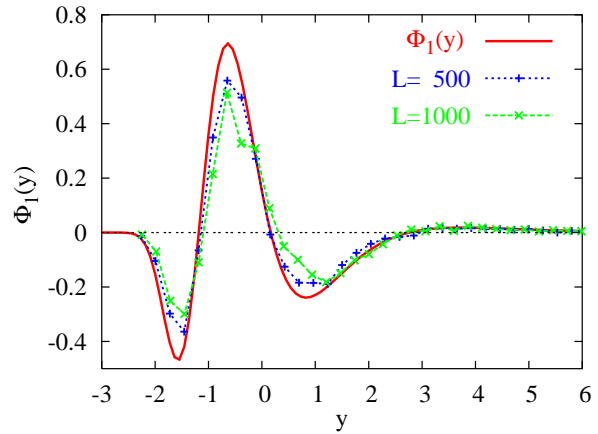}
\vspace{-10pt}
\caption{Finite-size analysis for the largest clusters in subcritical
percolation. The prediction from iid theory (solid line) is
$\Phi_1(y)=M^{(1)}(y)^\prime$ from Eq.~\eqref{eq:fs-ftg-null}.  The
simulation results (dotted line) were obtained with occupation
probability $p = 0.25$ and system sizes
$L=500,1000$. \label{F:fig_perc_fss}}
\vspace{-10pt}
\end{figure}

As an application, we studied the FS corrections to
the distribution of the largest cluster size on a
square lattice in subcritical site percolation, $p < p_c$, where
$p_c\approx0.592\dots$ is the critical occupation
probability.  Due to the finite
correlation length, clusters in a large system are nearly independent
and the size distribution of the largest obeys FTG, provided the
inherent discreteness of the problem is treated appropriately
\cite{Bazant:2000,HofstadRedig:2006}.  It remains an open question,
however, whether the FS corrections can also be described by iid theory.
To answer this question, we first note that the asymptote of the
distribution of the cluster size s is $s^{-1} \exp(-s/s_{\xi})$
\cite{StaufferAharony:1994}, where
$s_\xi$ is the cut-off size.  This asymptote
corresponds to $\delta=\theta=1$ in the example of the previous
paragraph, so \eqref{eq:fs-ftg-null} gives the FS scaling function and
$\epsilon_N\approx -1/\ln^2N$.  To compare theory with simulation, we
collected statistics for the largest cluster in systems of sizes
$L=500$ and $1000$ in an ensemble of $\approx 10^7$ runs, resulting in
a relatively smooth histogram. The shape correction was then obtained
by subtracting the empirical histogram from the FTG distribution,
multiplied by $\ln^2N$, where $N$ is the average number of
clusters. The result is compared with the iid theory in
Fig.~\ref{F:fig_perc_fss}.
As can be seen the curves match surprisingly well, suggesting that the
iid theory is also relevant for the FS corrections.

Next, we treat correlated time signals, $h(t)$, and study the FS
effects on the distribution of their maxima, $h_m$.  To have control
on correlations, $1/f^\alpha$ noise is chosen, where the Fourier
amplitudes are independent Gaussian variables, with variance
$f^{-\alpha}$, and uniform, random phase \cite{AntalETAL:2002}.  For
$\alpha=0$, the process is white noise, for $0<\alpha<1$ it is
stationary with correlations decaying as $t^{\alpha-1}$, while
for $\alpha>1$ the fluctuations of the signal diverge as
$t\to\infty$. The distribution of $h_m$ has recently been studied
intensively, see \cite{Burkhardt:2007} and references therein.  The
main features to be recalled here are that the FTG distribution
applies for $0\leq \alpha<1$ \cite{Berman:1964} while, for $\alpha>1$,
nontrivial distributions emerge whose shape depends on boundary
conditions and the reference point from where the maximum is
measured.  Here we concentrate on the FS correction of the
distribution of maxima, and to be specific, the maximum is measured
from the mean of a periodic signal.
\begin{figure}
%\vspace{5pt}
\includegraphics[width=8cm]{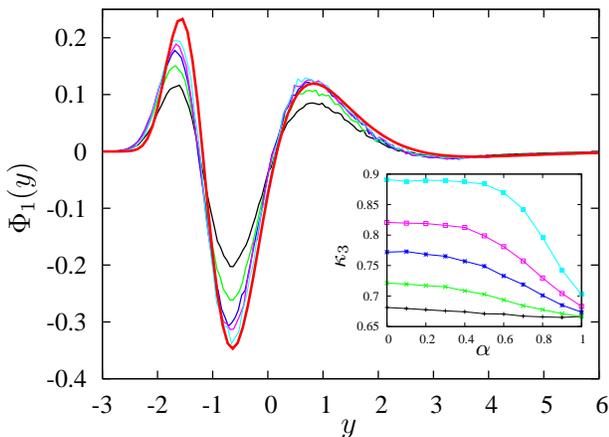}
\vspace{-10pt}
\caption{FS shape correction to the maximum distribution in
$1/f^\alpha$ noise with $\alpha=0.4$.  The solid line is the iid
theory $\Phi_1(y)=-M^{(1)}(y)^\prime/2$ from
Eq.~\eqref{eq:fs-ftg-null}, while the others are simulations for
system sizes $N = 2^5,2^7,2^9,2^{11},2^{15}$, with minimum decreasing in this order.  The inset shows the
skewness of the maximum distribution for various sizes slowly
converging to the FTG value $\kappa_3^0\approx 1.14$. It demonstrates
that FS properties may qualitatively change near $\alpha\approx 0.5$.
}\vspace{-10pt}
\label{F:fs_alpha_04}
\end{figure}
First, consider $\alpha<1$. Fig.~\ref{F:fs_alpha_04} shows the FS
shape correction for $\alpha=0.4$ together with the iid
prediction. The latter comes from a Gaussian parent, thus
$\epsilon_N\approx 1/2\ln N$, $\gamma'=0$. The theoretical curve is
$\Phi_1(y)=-M^{(1)}(y)^\prime/2$ by Eq.~\eqref{eq:fs-ftg-null},
corresponding to a $\ln{N}$ magnification factor in the simulation
curves which visibly approach $\Phi_1(x)$. A similar approach can be
seen for all $\alpha\lesssim 0.5$.  The inset, showing the skewness
for finite systems, also suggests that the leading FS correction may
be described by the iid theory for $\alpha\lesssim 0.5$.  This
conclusion goes beyond what we experienced in percolation: there
correlations had a finite cutoff, while here correlations decay like a
power.

%The present results are preliminary and obviously call for
%further investigations.

We now turn to $\alpha > 1$, where $\langle h_m\rangle$ diverges as
$\langle h_m\rangle \sim N^{(\alpha - 1)/2}$ \cite{GMOR:2007}, and the
approach to the limit distribution improves from logarithmic to
power-law. This effect can be seen in our simulations as well as
in the exact results \cite{MajumdarComtet:2004,MajumdarComtet:2005}
for random walks ($\alpha=2$). The limit distribution for $\alpha=2$
is given by the Airy distribution $\Phi_{Ai}(z)$ with $z=h_m/\sqrt{N}$
while the first correction to scaling is \cite{SchehrMajumdar:2006}
\begin{equation}
\Phi(z)\approx\Phi_{Ai}(z)+\Phi^\prime_{Ai}(z)/\sqrt{2N}
 \quad .
 \label{Majumdar-fss}
\end{equation}
The simplicity of the above result calls for simple explanation. Indeed,
Eq.~\eqref{Majumdar-fss} follows from the assumption that the
shape of the distribution relaxes faster than its position.  To see
this for arbitrary $\alpha>1$,
we note that the $n$-th cumulant of $h_m$ scales for large $N$
as $\kappa_n\sim N^{n\theta}$ with $\theta=(\alpha-1)/2$
\cite{GMOR:2007}.  Next, we write the corrections to scaling of
$\kappa_n$-s as
\begin{equation}
\kappa_n=N^{n\theta}(\kappa_n^{0}+\kappa_n^{1}N^{-\omega_n}+...)
 \, ,
 \label{kappa-expansion}
\end{equation}
and assume that $\omega_n>\omega_1$ for $n > 1$.  This assumption
implies that the shape of the function relaxes faster than its
position.  Introducing now the scaled variable $z=h_m/N^\theta$ and
expanding the cumulant generating function of $h_m$ in
$N^{-\omega_1}$ yields the scaled distribution function
$\Phi_N(z)=N^{\theta}P_N(N^{\theta}z)$ to first order as
\begin{equation}
\Phi_N(z)\approx\Phi(z)-\kappa_1^{1}\Phi^\prime (z)N^{-\omega_1}
 \label{Maj-fss-gen}
\end{equation}
where $\Phi(z)$ is the scaling function in the $N\to\infty$ limit.
For $\alpha=2$ one has $\omega_1=1/2$ and $\kappa_1^{1}=-1/\sqrt{2}$
\cite{SchehrMajumdar:2006}, thus Eq.~\eqref{Majumdar-fss} is recovered.
\begin{figure}[htb]
\vspace{-0pt}
\includegraphics[width=8.3cm]{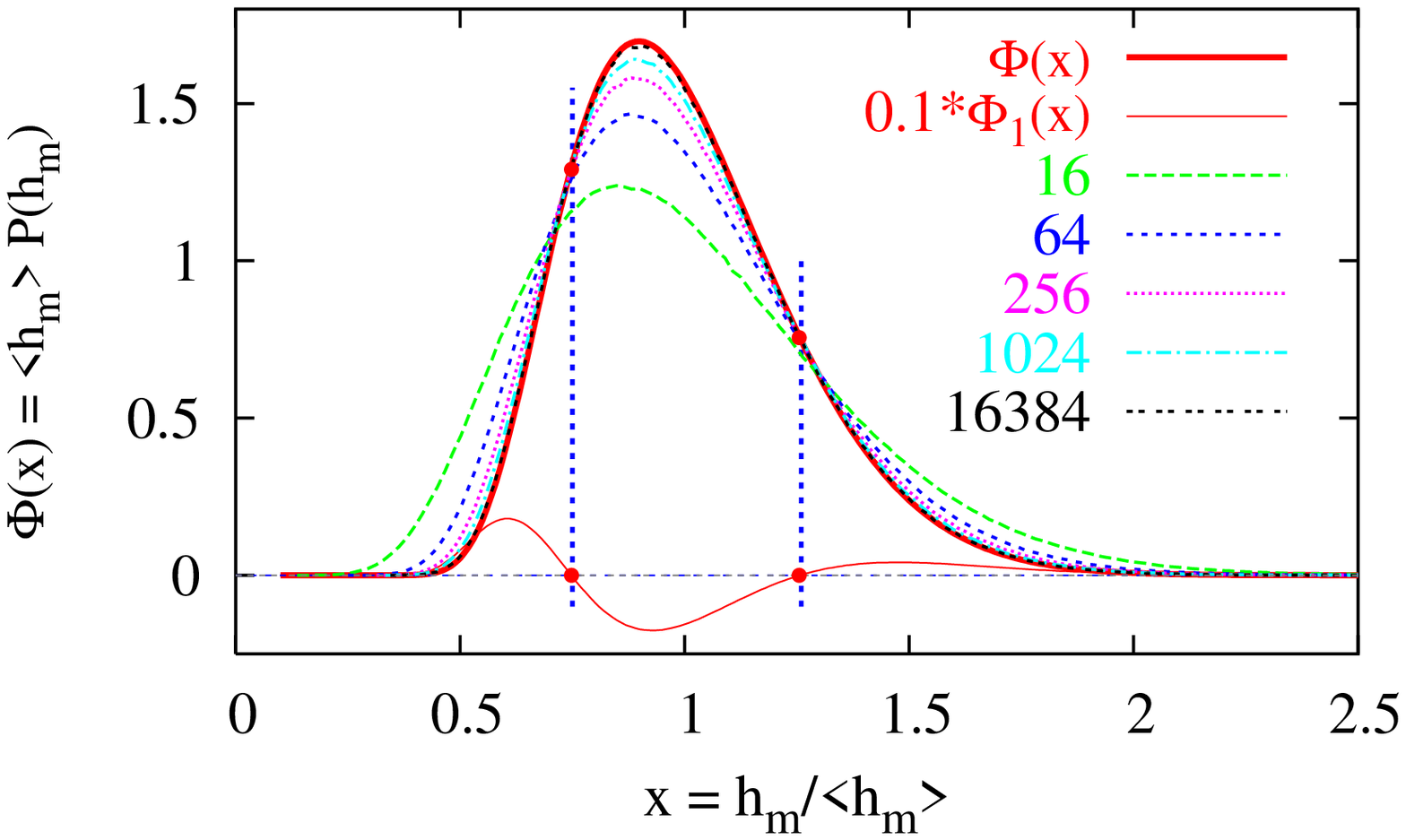}
\vspace{-0pt}
\includegraphics[width=8.3cm]{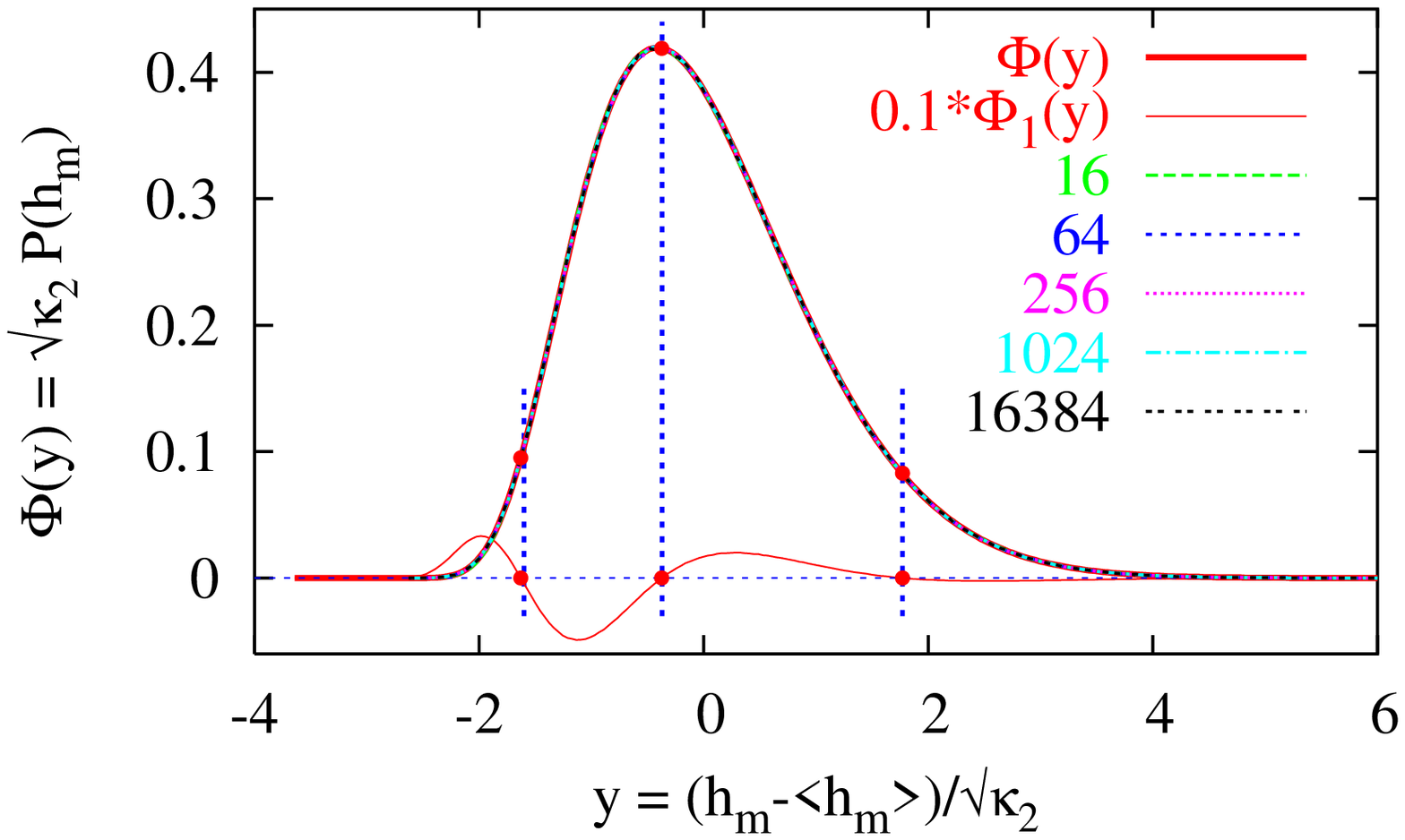}
\vspace{-0pt}
\caption{Demonstration of the increasing speed of convergence to the
limit distribution, $\Phi$, for $\alpha=2$. Results for system sizes
$N=16,\ldots,16384$ are shown using scaling variables $x=h_m/\langle
h_m\rangle $ and $y=(h_m - \langle h_m\rangle)/\sqrt{\kappa_2}$ in the
upper and lower panels, respectively.  $\Phi_1$ is the shape
correction function.
\label{F:a2.0-fss-comp}}
\end{figure}

The choice of scaling variable may change both the rate and amplitude
of the correction term. E.g. a natural choice is to scale by the
average [$x=h_m/\langle h_m \rangle$, $\Phi_N(x)=\langle h_m \rangle
P_N(\langle h_m \rangle x)$].  It yields the same rate of convergence
but it does alter the scaling function of the first order
correction. Indeed, using $x=h_m/\langle h_m \rangle$ and expanding
the cumulant generating function in $1/N^{\omega_1}$ results in
\begin{equation}
\Phi_N(x)\approx\Phi(x)-(\kappa_1^{1}/\kappa_1^{0})
\left[(x-1)\Phi(x)\right]^\prime N^{-\omega_1}
 \,.
 \label{fss-x-sc}
\end{equation}
The limiting function $\Phi(x)$ for $\alpha =2$ and its correction
$\Phi_1(x)$ with an amplitude $0.1$ are shown in the upper panel of
Fig.~\ref{F:a2.0-fss-comp}. Since $\Phi_1(x)$ has two zeros,
$\Phi_N(x)$ is nearly fixed at two points, so the main correction comes
from the central weight shifted to the tails of the distribution.

If the main FS effect is due to $\langle h_m\rangle$ then the use of
the scaling variable $y=(h_m-\langle h_m\rangle)/\sqrt{\kappa_2}$
eliminates these corrections and, as seen in Fig.~\ref{F:a2.0-fss-comp},
improves convergence dramatically.  The shape correction may be
calculated by assuming that $\omega_n>\omega_2$ for all $n>2$. This
means that the leading FS correction comes from $\kappa_2$ and the
cumulant generating function to first order in $1/N^{\omega_2}$ yields
\begin{equation}
{\Phi}_N(y)\approx \Phi(y)-(\kappa_2^{1}/2\kappa_2^{0})\,
[y\Phi(y)+\Phi^\prime(y)]^\prime\,N^{-\omega_2}\, .
 \label{fss-y-sc1}
\end{equation}
As seen, the shape correction can again be expressed in terms of the
limit distribution.  The scaling function $\Phi_1(y)$ displayed in
Fig.~\ref{F:a2.0-fss-comp} has three zeros which restrict possible
deviations from the limit distribution to higher order. In
addition, we found numerically $\omega_2 \approx 3/2$, resulting in
such a fast convergence that the curves with various $N$-s cannot be
distinguished within linewidth.

We can make an intuitive proposition for the $\omega_n$-s, based on an
analogy with the cumulants $\kappa_n(w_2)$ of the roughness $w_2$ of
$1/f^\alpha$ signals \cite{AntalETAL:2002}.  There, the large $N$
asymptote can be rewritten as $\kappa_n(w_2)\sim \left\langle w_2
\right\rangle^n (1-b_n\left\langle w_2
\right\rangle^{(1-n\alpha)/(\alpha -1)})$.  Now assuming the same
exponent $(1-n\alpha)/(\alpha -1)$ for $\kappa_n(h_m)$, and reverting
to the $N$ dependence we get $\omega_n(\alpha)=(n\alpha-1)/2$ for the
FS exponent.  For Brownian motion we recover
$\omega_2(2) = 3/2$, in accordance with simulations.  Furthermore, the
criterion $\omega_n<\omega_{n+1}$ is satisfied, so
$\omega_2(\alpha)=\alpha-1/2$ is the candidate for the FS
exponent. Since $\omega_n(\alpha)$ increases with $\alpha$, the
convergence is expected to improve for larger $\alpha$.  Indeed, our
simulations for $\alpha =4$ show that the same convergence as in the
lower panel of Fig.\ \ref{F:a2.0-fss-comp} can already be obtained in
the $x=h_m/\langle h_m\rangle$ scaling since
$\omega_1(4)=\omega_2(2)$.

So far we have considered the leading FS correction.
It is natural to ask about higher orders, especially in
the FTG class with typically slow, logarithmic convergence.
Higher order calculations are possible, which we just illustrate
here by an arbitrary order result for the parent
distribution $\mu(z)=1-e^{-z^\delta}$,
which, for  $\delta=2$, is the Rayleigh distribution, the basic
proposition for the statistics of wave crests
in ocean engineering \cite{Longuet-Higgins:1952}. For $\delta=2$,
the distribution of the maxima is obtained with
appropriate choice of $x=(z-b_N)/a_N$ as
\begin{eqnarray}
M_N(x) \approx \exp\left\lbrace -\exp \left[ -xH\left( \frac{x}{\ln
N}\right) \right]\right\rbrace+O(1/N),\label{eq:ray-fs}
\end{eqnarray}
with $ H(u)=[\left( 1+u/\delta\right)^\delta-1] /u$.
Remarkably, all logarithmic orders sum up to a scaling function
in the variable $x/\ln N$. For a general parent distribution,
the second order calculation has been carried out by a direct method
\cite{DeHaanResnick:1996}. Inspired by that, we have worked out an
algorithm to arbitrary orders which will be presented elsewhere.

Finally, we note that the RG approach can potentially be extended
to the study of EVS in correlated systems. In cases where the
limit distribution is known, the FS corrections may be clarified,
while in less explored systems it may help in finding the
limit distribution itself.

This work was supported by the Hungarian Academy of Sciences (Grant
No.\ OTKA K68109). NRM acknowledges support from the EU under a Marie
Curie Action.

%\bibliography{evs.bib}

\begin{thebibliography}{21}
\expandafter\ifx\csname natexlab\endcsname\relax\def\natexlab#1{#1}\fi
\expandafter\ifx\csname bibnamefont\endcsname\relax
  \def\bibnamefont#1{#1}\fi
\expandafter\ifx\csname bibfnamefont\endcsname\relax
  \def\bibfnamefont#1{#1}\fi
\expandafter\ifx\csname citenamefont\endcsname\relax
  \def\citenamefont#1{#1}\fi
\expandafter\ifx\csname url\endcsname\relax
  \def\url#1{\texttt{#1}}\fi
\expandafter\ifx\csname urlprefix\endcsname\relax\def\urlprefix{URL }\fi
\providecommand{\bibinfo}[2]{#2}
\providecommand{\eprint}[2][]{\url{#2}}

\bibitem[{\citenamefont{Gumbel}(1958)}]{Gumbel:1958}
\bibinfo{author}{\bibfnamefont{E.~J.} \bibnamefont{Gumbel}},
  \emph{\bibinfo{title}{Statistics of Extremes}} (\bibinfo{publisher}{Dover},
  \bibinfo{year}{1958}).

\bibitem[{\citenamefont{Embrecht et~al.}(1997)\citenamefont{Embrecht,
  Kl\"uppelberg, and Mikosch}}]{EmrechtETAL:2004}
\bibinfo{author}{\bibfnamefont{P.}~\bibnamefont{Embrecht}},
  \bibinfo{author}{\bibfnamefont{C.}~\bibnamefont{Kl\"uppelberg}},
  \bibnamefont{and} \bibinfo{author}{\bibfnamefont{T.}~\bibnamefont{Mikosch}},
  \emph{\bibinfo{title}{Modelling Extremal Events for Insurance and Finance}}
  (\bibinfo{publisher}{Springer, Berlin}, \bibinfo{year}{1997}).

\bibitem[{\citenamefont{Katz et~al.}(2002)\citenamefont{Katz, Parlange, and
  Naveau}}]{KatzParlangeNaveau:2002}
\bibinfo{author}{\bibfnamefont{R.~W.} \bibnamefont{Katz}},
  \bibinfo{author}{\bibfnamefont{M.~B.} \bibnamefont{Parlange}},
  \bibnamefont{and} \bibinfo{author}{\bibfnamefont{P.}~\bibnamefont{Naveau}},
  \bibinfo{journal}{Adv.~Water~Resour.} \textbf{\bibinfo{volume}{25}},
  \bibinfo{pages}{1287} (\bibinfo{year}{2002}).

\bibitem[{\citenamefont{Bouchaud and M\'ezard}(1997)}]{MezardETAL:1997}
\bibinfo{author}{\bibfnamefont{J.-P.} \bibnamefont{Bouchaud}} \bibnamefont{and}
  \bibinfo{author}{\bibfnamefont{M.}~\bibnamefont{M\'ezard}},
  \bibinfo{journal}{JPA} \textbf{\bibinfo{volume}{30}}, \bibinfo{pages}{7997}
  (\bibinfo{year}{1997}).

\bibitem[{\citenamefont{Gy\"{o}rgyi et~al.}(2003)\citenamefont{Gy\"{o}rgyi,
  Holdsworth, Portelli, and R\'{a}cz}}]{GyorgyiETAL:2003}
\bibinfo{author}{\bibfnamefont{G.}~\bibnamefont{Gy\"{o}rgyi}},
  \bibinfo{author}{\bibfnamefont{P.}~\bibnamefont{Holdsworth}},
  \bibinfo{author}{\bibfnamefont{B.}~\bibnamefont{Portelli}}, \bibnamefont{and}
  \bibinfo{author}{\bibfnamefont{Z.}~\bibnamefont{R\'{a}cz}},
  \bibinfo{journal}{Phys. Rev. E} \textbf{\bibinfo{volume}{68}},
  \bibinfo{pages}{056116} (\bibinfo{year}{2003}).

\bibitem[{\citenamefont{Majumdar and Comtet}(2004)}]{MajumdarComtet:2004}
\bibinfo{author}{\bibfnamefont{S.}~\bibnamefont{Majumdar}} \bibnamefont{and}
  \bibinfo{author}{\bibfnamefont{A.}~\bibnamefont{Comtet}},
  \bibinfo{journal}{Phys. Rev. Lett.} \textbf{\bibinfo{volume}{92}},
  \bibinfo{pages}{225501} (\bibinfo{year}{2004}).

\bibitem[{\citenamefont{Krapivsky and Majumdar}(2000)}]{KrapivskyMajumdar:2000}
\bibinfo{author}{\bibfnamefont{P.}~\bibnamefont{Krapivsky}} \bibnamefont{and}
  \bibinfo{author}{\bibfnamefont{S.}~\bibnamefont{Majumdar}},
  \bibinfo{journal}{Phys. Rev. Lett.} \textbf{\bibinfo{volume}{85}},
  \bibinfo{pages}{5492} (\bibinfo{year}{2000}).

\bibitem[{\citenamefont{de~Haan and Ferreira}(2006)}]{DeHaanFerreira:2006}
\bibinfo{author}{\bibfnamefont{L.}~\bibnamefont{de~Haan}} \bibnamefont{and}
  \bibinfo{author}{\bibfnamefont{A.}~\bibnamefont{Ferreira}},
  \emph{\bibinfo{title}{Extreme {V}alue {T}heory: {A}n {I}ntroduction}}
  (\bibinfo{publisher}{Springer, New York}, \bibinfo{year}{2006}).

\bibitem[{\citenamefont{Schehr and Majumdar}(2006)}]{SchehrMajumdar:2006}
\bibinfo{author}{\bibfnamefont{G.}~\bibnamefont{Schehr}} \bibnamefont{and}
  \bibinfo{author}{\bibfnamefont{S.~N.} \bibnamefont{Majumdar}},
  \bibinfo{journal}{Phys. Rev. E} \textbf{\bibinfo{volume}{73}},
  \bibinfo{pages}{056103} (\bibinfo{year}{2006}).

\bibitem[{\citenamefont{Bazant}(2000)}]{Bazant:2000}
\bibinfo{author}{\bibfnamefont{M.~Z.} \bibnamefont{Bazant}},
  \bibinfo{journal}{Phys. Rev. E} \textbf{\bibinfo{volume}{62}},
  \bibinfo{pages}{1660} (\bibinfo{year}{2000}).

\bibitem[{\citenamefont{van~der Hofstad and Redig}(2006)}]{HofstadRedig:2006}
\bibinfo{author}{\bibfnamefont{R.}~\bibnamefont{van~der Hofstad}}
  \bibnamefont{and} \bibinfo{author}{\bibfnamefont{F.}~\bibnamefont{Redig}},
  \bibinfo{journal}{J. Stat. Phys.} \textbf{\bibinfo{volume}{122}},
  \bibinfo{pages}{671} (\bibinfo{year}{2006}).

\bibitem[{\citenamefont{Berman}(1964)}]{Berman:1964}
\bibinfo{author}{\bibfnamefont{S.~M.} \bibnamefont{Berman}},
  \bibinfo{journal}{Ann.~Math.~Statist.} \textbf{\bibinfo{volume}{33}},
  \bibinfo{pages}{502} (\bibinfo{year}{1964}).

\bibitem[{\citenamefont{Galambos}(1978)}]{Galambos:1978}
\bibinfo{author}{\bibfnamefont{J.}~\bibnamefont{Galambos}},
  \emph{\bibinfo{title}{The {A}symptotic {T}heory of {E}xtreme {V}alue
  {S}tatistics}} (\bibinfo{publisher}{John Wiley \& Sons},
  \bibinfo{year}{1978}).

\bibitem[{\citenamefont{Fisher and Tippett}(1928)}]{FisherTippett:1928}
\bibinfo{author}{\bibfnamefont{R.}~\bibnamefont{Fisher}} \bibnamefont{and}
  \bibinfo{author}{\bibfnamefont{L.}~\bibnamefont{Tippett}},
  \bibinfo{journal}{Procs.~Cambridge~Philos.~Soc.}
  \textbf{\bibinfo{volume}{24}}, \bibinfo{pages}{180} (\bibinfo{year}{1928}).

\bibitem[{\citenamefont{de~Haan and Resnick}(1996)}]{DeHaanResnick:1996}
\bibinfo{author}{\bibfnamefont{L.}~\bibnamefont{de~Haan}} \bibnamefont{and}
  \bibinfo{author}{\bibfnamefont{S.}~\bibnamefont{Resnick}},
  \bibinfo{journal}{Annals of Prob.} \textbf{\bibinfo{volume}{24}},
  \bibinfo{pages}{97} (\bibinfo{year}{1996}).

\bibitem[{\citenamefont{Stauffer and Aharony}(1994)}]{StaufferAharony:1994}
\bibinfo{author}{\bibfnamefont{D.}~\bibnamefont{Stauffer}} \bibnamefont{and}
  \bibinfo{author}{\bibfnamefont{A.}~\bibnamefont{Aharony}},
  \emph{\bibinfo{title}{Introduction To Percolation Theory}}
  (\bibinfo{publisher}{Taylor and Francis, London}, \bibinfo{year}{1994}).

\bibitem[{\citenamefont{Antal et~al.}(2002)\citenamefont{Antal, Droz,
  Gy\"{o}rgyi, and R\'{a}cz}}]{AntalETAL:2002}
\bibinfo{author}{\bibfnamefont{T.}~\bibnamefont{Antal}},
  \bibinfo{author}{\bibfnamefont{M.}~\bibnamefont{Droz}},
  \bibinfo{author}{\bibfnamefont{G.}~\bibnamefont{Gy\"{o}rgyi}},
  \bibnamefont{and} \bibinfo{author}{\bibfnamefont{Z.}~\bibnamefont{R\'{a}cz}},
  \bibinfo{journal}{Phys. Rev. E} \textbf{\bibinfo{volume}{65}},
  \bibinfo{pages}{046140} (\bibinfo{year}{2002}).

\bibitem[{\citenamefont{Burkhardt et~al.}(2007)\citenamefont{Burkhardt,
  Gy\"{o}rgyi, Moloney, and R\'{a}cz}}]{Burkhardt:2007}
\bibinfo{author}{\bibfnamefont{T.~W.} \bibnamefont{Burkhardt}},
  \bibinfo{author}{\bibfnamefont{G.}~\bibnamefont{Gy\"{o}rgyi}},
  \bibinfo{author}{\bibfnamefont{N.~R.} \bibnamefont{Moloney}},
  \bibnamefont{and} \bibinfo{author}{\bibfnamefont{Z.}~\bibnamefont{R\'{a}cz}},
  \bibinfo{journal}{Phys. Rev. E} \textbf{\bibinfo{volume}{76}},
  \bibinfo{pages}{041119} (\bibinfo{year}{2007}).

\bibitem[{\citenamefont{Gy\"{o}rgyi et~al.}(2007)\citenamefont{Gy\"{o}rgyi,
  Moloney, Ozog\'{a}ny, and R\'{a}cz}}]{GMOR:2007}
\bibinfo{author}{\bibfnamefont{G.}~\bibnamefont{Gy\"{o}rgyi}},
  \bibinfo{author}{\bibfnamefont{N.~R.} \bibnamefont{Moloney}},
  \bibinfo{author}{\bibfnamefont{K.}~\bibnamefont{Ozog\'{a}ny}},
  \bibnamefont{and} \bibinfo{author}{\bibfnamefont{Z.}~\bibnamefont{R\'{a}cz}},
  \bibinfo{journal}{Phys. Rev. E} \textbf{\bibinfo{volume}{75}},
  \bibinfo{pages}{021123} (\bibinfo{year}{2007}).

\bibitem[{\citenamefont{Majumdar and Comtet}(2005)}]{MajumdarComtet:2005}
\bibinfo{author}{\bibfnamefont{S.}~\bibnamefont{Majumdar}} \bibnamefont{and}
  \bibinfo{author}{\bibfnamefont{A.}~\bibnamefont{Comtet}},
  \bibinfo{journal}{J.~Stat.~Phys.} \textbf{\bibinfo{volume}{119}},
  \bibinfo{pages}{777} (\bibinfo{year}{2005}).

\bibitem[{\citenamefont{Longuet-Higgins}(1952)}]{Longuet-Higgins:1952}
\bibinfo{author}{\bibfnamefont{M.~S.} \bibnamefont{Longuet-Higgins}},
  \bibinfo{journal}{J. Mar. Res.} \textbf{\bibinfo{volume}{11}},
  \bibinfo{pages}{245} (\bibinfo{year}{1952}).

\end{thebibliography}

\end{document}